\documentclass[10pt]{article}
\usepackage{amsmath, amssymb}

\usepackage{geometry}
\usepackage{float}
\usepackage{paralist} 
\usepackage{graphics}
\usepackage{subfigure}
\usepackage{epsfig}
\usepackage{graphicx}
\usepackage{threeparttable}
\usepackage[colorlinks,linkcolor=black,anchorcolor=black,citecolor=black]{hyperref}
\usepackage[sort&compress,numbers]{natbib} 
\usepackage{soul}

\newcommand\Tstrut{\rule{0pt}{2.6ex}}         
\newcommand\Bstrut{\rule[-1.3ex]{0pt}{0pt}}   

\newcommand\norm[1]{\lVert#1\rVert}

\title{EISA-Score: Element Interactive Surface Area Score for Protein-Ligand Binding Affinity Prediction}
 
\author{Md Masud Rana$^1$ and Duc Duy Nguyen$^1$ \footnote{Address correspondences to Duc Duy Nguyen. E-mail: ducnguyen@uky.edu}\\
$^1$ Department of Mathematics,
University of Kentucky, KY 40506, USA\\
}
\begin{document}

\maketitle

\begin{abstract}
Molecular surface representations have been advertised as a great tool to study protein structure and functions, including protein-ligand binding affinity modeling. However, the conventional surface-area-based methods fail to deliver a competitive performance on the energy scoring tasks. The main reason is the lack of crucial physical and chemical interactions encoded in the molecular surface generations. We present novel molecular surface representations embedded in different scales of the element interactive manifolds featuring the dramatically dimensional reduction and accurately physical and biological properties encoders. Those low-dimensional surface-based descriptors are ready to be paired with any advanced machine learning algorithms to explore the essential structure-activity relationships that give rise to the element interactive surface area-based scoring functions (EISA-score). The newly developed EISA-score has outperformed many state-of-the-art models, including various well-established surface-related representations, in standard PDBbind benchmarks.

\end{abstract}

\section{Introduction}
Geometric modeling of biomolecules concerns geometrical components at various scales and dimensions, including molecular surface generation, molecular visualization, curvature analysis, surface annotation, etc. \cite{corey1953molecular,koltun1965precision,lee1971interpretation,richards1977areas,connolly1985depth,grant2001smooth, vorobjev1997sims,  sanner1996reduced, grant2007gaussian, li2013dielectric, wang2015pka}. Among these ingredients, the molecular surface plays a significant role in visualizing and analyzing molecular structures and properties. Specifically, one can project the electrostatic potentials, flexibility indexes, and curvature magnitudes on the protein surface \cite{petrey2003grasp2} to reveal protein structure and function, such as protein-ligand binding sites, protein-protein binding hot spots, and protein-DNA interactions.

There are various methods proposed to compute the biomolecular surfaces. One can classify these methods into three categories: analytical representation, partial differential equations (PDEs)-based generation, and explicit formulation. For the analytical calculation, the simplest model can be referred to van der Waals surface (vdWS), formed by a union of the atomic sphere of the Van der Waals radius. In addition, one can use the trajectory of the probe's center moving around the van der Waals surface to give rise to the Solvent accessible surface (SAS) \cite{lee1971interpretation}. Unfortunately, those vdWS and SAS approaches suffer the non-smooth regions causing computation obstacles. For that reason, Connolly proposed a solvent excluded surface (SES) to avoid these non-smooth issues \cite{connolly1985depth}. MSMS software was later developed to improve the speed and reliability of SES calculation via the reduced surface \cite{sanner1996reduced}. There are other efficient algorithms for generating SES \cite{chen2010kernel,chan1998molecular,edelsbrunner1994three,fraczkiewicz1998exact,hayryan2005new,liang1998analytical,rychkov2007joint,xu2009generating, lange2020intrinsically,daberdaku2018computing, egan2018fast,hermosilla2017interactive, liu2018efficient,liu2017eses}. Among them, TMSmesh \cite{liu2018efficient} used boundary element method and finite element method to handle arbitrary sizes of molecules.
By adapting a multi-step region-growing EDT approach, Daberdaku and Ferrari \cite{daberdaku2018computing} developed fast molecular surface representations for large molecules. Hermosilla et al. \cite{hermosilla2017interactive} utilized interactive GPU power to accelerate SES rendering at a fractional cost. Wei and his co-workers  introduced ESES which accurately generates SES on the Cartesian mesh \cite{liu2017eses}.

To define the solute-solvent region, one can allow the overlap of the solvent and solute domains via the fuzzy characteristic or hypersurface functions. This approach was initially introduced in 2005 to generate the class of desirable biomolecular surfaces by curvature-driven geometric PDEs \cite{wei2005molecular}. The other type used the mean curvature flow or Laplace-Beltrami equation to form the molecular surface by minimizing the surface energy \cite{bates2006minimal,bates2008minimal,bates2009geometric}. Later, these optimal geometric flow schemes have been extended to model the nonpolar energy of biomolecular systems \cite{wei2012variational,wei2013multiscale,chen2010differential,chen2011differential,chen2011differential2}.

Despite the fact that the analytical approaches are able to generate the accurate molecular surface and PDE-based methods can embed the molecular energy information, they are not flexible when expressing the surface of local atoms. In addition, the analytical surfaces often consist of geometric singularities which obstruct the estimation of other geometry information such as curvature \cite{connolly1985depth, sanner1996reduced}. The prominent representative of the explicit surface are the Gaussian surfaces, in which the Gaussian functions are used as the density potential at each atom \cite{grant1995gaussian,yu2008feature,mu2017geometric,nguyen2017impact}. Those surfaces avoid the geometric singularities but are sensitive to level set values used to extract a specific surface candidate \cite{mu2017geometric}.

Molecular surface representations have shown their important role in predictions of solvation-free energies and ion channel transport. However, they have to be incorporated in the realm of the physical models such as Poisson-Boltzmann equation \cite{chen2012variational, wang2015parameter} and Poisson-Nernst-Planck model \cite{chen2012quantum, chen2012quantum_2, wei2012variational, wei2013multiscale, chen2013quantum}. These dependencies limit the direct link of the molecular surface properties on the molecular properties. In addition, the dependence on the parameterized factors such as atomic charges and grid sizes of the predefined domain has restrained the capability of the molecular surface details on the diverse and complex biomolecular structures \cite{nguyen2017accurate}.

Due to the essential physical and chemical properties captured on the molecular surface, its information has been widely used in quantitative and qualitative tasks in exploring molecule properties and activities. For the qualitative purpose, the biomolecular surface can be used to visualize  protein folding \cite{spolar1994coupling}, protein-protein interactions \cite{crowley2005cation}, DNA binding and bending \cite{dragan2004dna}, molecular docking \cite{sobolev1996molecular}, binding site classification \cite{das2009rapid}, and molecular dynamic \cite{geng2011multiscale}. In the quantitative effort, the molecular surface can be integrated with the implicit solvent model to prediction solvation free energy \cite{baker2005improving, chen2011mibpb, nguyen2017impact}, incorporated with the Poisson–Nernst–Planck setting to compute the electrostatic and concentration profiles, and current-voltage curves \cite{zheng2011second, zheng2011poisson}, used as a variable in the partial least squares model to predict the solubility and permeability of the druglike molecules \cite{bergstrom2003absorption}. However, those approaches are limited in representing complex biomolecular structures from large and diverse datasets due to the lack of details of physical and chemical interactions.

Recently, we have unlocked the representation power of the curvatures of the molecular surface for massive and distinct molecular and biomolecular structures to predict drug toxicity, molecular solvation energy, and protein-ligand binding affinity \cite{nguyen2019dg}. However, the role of the surface area in capturing the crucial physical and chemical interactions in the biomolecular structures is not fully explored. Despite the recent efforts to integrate the surface area information into the predictive models such as Cyscore \cite{cao2014improved} and GLXE \cite{dong2021prediction} for protein-ligand binding affinity prediction, those surface area-based models are far from the competitive level with their counterparts. 

To decipher the full potential of the surface area-based descriptors, we propose to construct the molecular surface at the pairwise element levels. The element-wise surfaces will effectively capture some specific types of non-covalent interactions, such as Van der Waals interactions, hydrophobicity, and hydrogen bonds. Furthermore, the element-level surface area features highlight the scalability in the sense that the proposed representation will be independent of molecular sizes, i.e., number of atoms, thus enabling the equal footing configuration for molecular structures from the highly diversified datasets. Given the information of atomic coordinates, there are several ways to construct the corresponding molecular surface. In this work, we extend our proposed molecular surface generation of small molecules in the implicit solvation modeling \cite{nguyen2017impact} to characterize the surfaces between protein and ligand at the element level. In general, the Riemannian manifolds are constructed on the subsets of the group of element types to allow to conveniently form the structures of differential geometry. One can extract the manifold representations for the selected atoms via a discrete-to-continuum mapping that enables the embedding of the high dimensional data space of the biomolecular atoms into the low-dimensional model \cite{xia2013multiscale,xia2015multiresolution,xia2016review}.

The objective of the present work is to introduce the element-interactive surface area (EISA) descriptors for the first time in the literature to accurately and effectively describe the molecular representations in the low-dimensional space. The interactive molecular surface is presented by the standard correlation functions, namely exponential and Lorentz kernel functions which give us the Gaussian-like surfaces. Moreover, those surfaces are infinitely differentiable and free of geometric singularities. In this work, we are interested in constructing a class of surfaces at the multiscale levels by varying the suitable kernel parameters and level set values via the multiscale discrete-to-continuum mapping. By pairing with the advanced machine learning architectures, the molecular surface-based model, named EISA-Score, reveal its quantitative power in predicted drug-related molecular properties, such as protein-ligand binding affinity (BA). The accurate and robust method to calculate the BA values of the small molecules is the crucial component in speeding up the process of drug discovery to help design novel drugs. In this work, we testify the scoring power of our proposed model against three commonly benchmarks in drug design area, namely CASF-2007 \cite{cheng2009comparative}, CASF-2013 \cite{li2014comparative}, and CASF-2016 \cite{su2018comparative}. Several experiments confirm that our EISA-Score achieves state-of-the-art results and outperforms the other molecular surface-based models by a wide margin.

\section{Model Development}

\subsection{Element interactive manifolds}
This section presents a background of the discrete-to-continuum mapping via the atomic density function formulated in the common choice of correlation kernel functions. Under the element-wise setting, that mapping extracts the low-dimensional manifolds targeting the specific element types to represent the high dimensional interactions for the group of atoms of interest.

\subsubsection{Atomic density}
Given a molecule with $N$ atoms, we denote $\mathcal{X} = \{\mathbf{r}_1,\mathbf{r}_2,\cdots, \mathbf{r}_N \}$ the set of $N$ atomic coordinates. Let $\mathbf{r}_j\in \mathbb{R}^3$ be the position of $j$th atom in the molecule and $\lVert \mathbf{r}-\mathbf{r}_j \rVert$ be the Euclidean distance between the atom $\mathbf{r}_j$ and a point $\mathbf{r}\in \mathbb{R}^3$. The molecular density is given by a discrete-to-continuum mapping
\begin{align}\label{eq:atomic_density}
    \rho(\mathbf{r})= \sum_{j=1}^N \omega_j \Phi(\norm{\mathbf{r}-\mathbf{r}_j};\eta_j),
\end{align}
where $\omega_j$ are the weights, $\eta_j$ are characteristic distances, and $\Phi$ is a $C^2$ correlation kernel or statistical density estimator that satisfies the following admissibility conditions

\begin{align}
    \Phi(\norm{\mathbf{r}-\mathbf{r}_j};\eta_j) &=1, \quad \mathrm{as}\; \norm{\mathbf{r}-\mathbf{r}_j}\rightarrow 0,\\
    \Phi(\norm{\mathbf{r}-\mathbf{r}_j};\eta_j) &=0, \quad \mathrm{as}\; \norm{\mathbf{r}-\mathbf{r}_j}\rightarrow \infty.
\end{align}

As in our previous work \cite{nguyen2017rigidity,nguyen2019dg, nguyen2019agl}, the generalized exponential and generalized Lorentz functions have shown their robustness and efficiency in capturing the dynamic interactions between various types of atoms at different ranges. Their formulations are givens as the following
\begin{equation}
    \Phi_E(\norm{\mathbf{r}-\mathbf{r}_j};\eta_j) = e^{-(\norm{\mathbf{r}-\mathbf{r}_j}/\eta_j)^\kappa}, \quad \kappa>0; \quad\text{(generalized exponential kernel)}
\end{equation}

\begin{equation}
    \Phi_L(\norm{\mathbf{r}-\mathbf{r}_j};\eta_j) = \frac{1}{1+\left(\norm{\mathbf{r}-\mathbf{r}_j}/\eta_j\right)^\kappa}, \quad \kappa>0 \quad \text{(generalized Lorentz kernel)}.
\end{equation}
In the present work, the atomic weights $w_j$ are chosen to be 1 for simplicity. In other applications, one might consider the atomic charges to represent the atomic weights \cite{nguyen2019dg,cang2018representability}. The kernel parameters $\eta_j$ and $\kappa$ need to be carefully selected to capture the crucial interactions between different atom types and consequently produce a meaningful molecular surface. The multiscale atom density that can be obtained by choosing different ranges for the kernel parameter sets $\eta_j$, and $\kappa$ has shown its potential in covering different wide range intramolecular interactions from the diverse families of proteins \cite{nguyen2017rigidity,opron2015communication}.

\subsubsection{Element interactive densities}
To account for details of physical interactions in protein-ligand complex such as hydrophobic, hydrophilic, etc., we are interested in constructing the atomic densities in an element interactive manner. To this end, we consider the four most appearances element types in protein, namely C, N, O, and S, while there are ten commonly occurring element types in ligand, namely H, C, N, O, S, P, F, Cl, Br, and I. As a result, we have 40 element interactive possibilities between protein and ligand atoms: HH, HC, HO, …, HI, CH, …, and SI. Although our discussing element specifics are designed for the protein-ligand system, this approach, with minimal effort, can be applied to a single biomolecular setting and other interactive models in chemistry and biology.

For convenience, we let $\mathcal{T}=\{\mathrm{H, C, N, O, S, P, F, Cl, \dots}\}$ is the set of all our interested element types in a given biomolecular dataset. To reduce the notation complexity, we denote the element type at the $i$th position in the set $\mathcal{T}$ as $\mathcal{T}_i$. For example, $\mathcal{T}_2$ indicates the element type Carbon. Assuming that a biomolecule has $N$ atoms of interest. Then, we assign $\mathcal{X}=\{ (\mathbf{r}_i,\alpha_i)| \mathbf{r}_i\in \mathbb{R}^3; \alpha_i\in \mathcal{T}; i=1,2,\cdots,N\}$ as the collection of these $N$ atoms annotated by their coordinates $\mathbf{r}_i$ and element types $\alpha_i$. Before constructing the element interactive densities, we define the element interactive domain $D_{kk'}$ for element type $\mathcal{T}_k$ and $\mathcal{T}_{k'}$ as the following

\begin{equation}
    D_{kk'} = \left\{\mathbf{r}\in\bigcup_i B(\mathbf{r}_i, d_c) \vert \alpha_i \in \{\mathcal{T}_k, \mathcal{T}_{k'}\}, i=1,2,\dots,N\right\},
\end{equation}
where $d_c$ is a predefined cut-off distance, and $B(\mathbf{r}_i, d_c)$ is a ball with a center $\mathbf{r}_i$ and a radius $d_c$. We now can design the element interactive density $\rho_{kk'}$, an atomic density defined in (\ref{eq:atomic_density}) but with a restraint on the element interactive region $D_{kk'}$:

\begin{equation}\label{eq:global_element_specific_density}
    \rho_{kk'}(\mathbf{r},\Phi) =  \sum_j \Phi(\norm{\mathbf{r}-\mathbf{r}_j};\eta_{kk'}), \quad \mathbf{r}\in D_{kk'}, \alpha_j\in\{\mathcal{T}_k, \mathcal{T}_{k'}\}.
\end{equation}

and the normalized density function can be defined as
\begin{align}
    \hat{\rho}_{kk'}(\mathbf{r}, \Phi) = \frac{\rho_{kk'}(\mathbf{r}, \Phi)}{\mathrm{max}\{\rho_{kk'}(\mathbf{r}, \Phi)\}}.
\end{align}

In this work, we call the density function (\ref{eq:global_element_specific_density}) the ``global density'' for the element types $\mathcal{T}_k$ and $\mathcal{T}_{k'}$. In addition, we desire to explore the ``local density'' formed by a single atom $r_{i_o}$ with an element type $\mathcal{T}_k$ and all element type $\mathcal{T}_{k'}$ atoms:
\begin{equation}\label{eq:local_element_specific_density}
    \rho_{kk'}^{i_o}(\mathbf{r},\Phi) =  \Phi(\norm{\mathbf{r}-\mathbf{r}_{i_o}};\eta_{kk'}) + \sum_{j\neq i_o} \Phi(\norm{\mathbf{r}-\mathbf{r}_j};\eta_{kk'}), \quad \mathbf{r}\in D^{i_o}_{kk'}, \alpha_j= \mathcal{T}_{k'},
\end{equation}
where $ D^{i_o}_{kk'}$ is a local element interactive domain defined as
\begin{equation}
    D^{i_o}_{kk'} = \left\{\mathbf{r}\in\bigcup_{j\neq i_o} B(\mathbf{r}_j, d_c) \cup B(\mathbf{r}_{i_o}, d_c) \vert \alpha_j =\mathcal{T}_{k'}, j=1,2,\dots,N\right\},
\end{equation}
and it is straightforward to verify that
\begin{equation}
    D_{kk'} = \bigcup_{i_o} D^{i_o}_{kk'}, \quad i_o=1,2,\dots,N \text{~and~} \alpha_{i_o} = \mathcal{T}_k.
\end{equation}

The assembly of the local element interactive density $\rho_{kk'}^{i_o}$ enables our proposed model to examine the local interactions between a single atom of the element type $\mathcal{T}_k$ against a group of atoms of the element type $\mathcal{T}_{k'}$, capturing essential physical and chemical information across different biomolecular families that the global density might omit. 
\subsection{Element interactive surface area}
With $\rho$ being a level set function defined on every grid point in an interested domain $\Omega$, the isosurface $\Gamma$ induced by $\rho$ is given by $\Gamma=\{(x,y,z)\in \Omega:\rho(x,y,z)=c\}$, where $c$ is the recommended isovalue. Assume $f(x,y,z)$ is the surface density function defined in $\Gamma$, the surface integral of $f$ in Cartesian grids with a uniform mesh can be evaluated by \cite{smereka2006numerical,geng2011multiscale}

\begin{equation}\label{surface-area}
    \int_{\Gamma} f(x,y,z)\; \mathrm{d}S = \sum_{(i,j,k)\in I_o} \left(f(x_o,y_j,z_k)\frac{|n_{o,x}|}{h} + f(x_i,y_o,z_k)\frac{|n_{o,y}|}{h} + f(x_i,y_j,z_o)\frac{|n_{o,z}|}{h} \right)h^3,
\end{equation}

where $h$ is the mesh size, $(x_o,y_j,z_k)$ is the intersection point between the interface $\Gamma$ and the $x$ mesh line going through $(i,j,k)$, and $n_{o,x}$ is the $x$ component of the unit normal vector at $(x_o,y_j,z_k)$. Similar definitions are used for the $y$ and $z$ directions. In addition, $I_o$ is the set of irregular grid points. In our numerical scheme, a grid point is classified as irregular if its numerical difference's stencil involves neighbor point(s) from the other side of the interface $\Gamma$.
One can find the surface area of $\Gamma$ by considering the density function $f=1$ in equation \eqref{surface-area}.


The intersection point $(x_o,y_j,z_k)$ can be determined as described in \cite{mu2017geometric} by
\begin{equation}
    (x_o,y_j,z_k) = \left(\frac{\rho(x_o,y_j,z_k)-\rho(x_i,y_j,z_k)}{\rho(x_{i+1},y_j,z_k)-\rho(x_i,y_j,z_k)}(x_{i+1}-x_i),y_j,z_k\right),
\end{equation}
where $\rho(x_o,y_j,z_k)=c$, and the corresponding normal vector at $(x_o,y_j,z_k)$ is interpolated by 

\begin{equation}\label{eqn:normal_vector}
    \mathbf{N}_{o,j,k} = \frac{\rho(x_o,y_j,z_k)-\rho(x_i,y_j,z_k)}{\rho(x_{i+1},y_j,z_k)-\rho(x_i,y_j,z_k)}\left(\mathbf{N}_{i+1,j,k}-\mathbf{N}_{i,j,k} \right) + \mathbf{N}_{i,j,k},
\end{equation}
where $\mathbf{N}_{i,j,k}$ is the normal vector at the grid point $(i,j,k)$ and is approximated by 
\begin{align*}
    \mathbf{N}_{i,j,k} = &\left(\frac{\rho(x_{i+1},y_j,z_k)-\rho(x_{i-1},y_j,z_k)}{x_{i+1}-x_{i-1}}, \frac{\rho(x_{i},y_{j+1},z_k)-\rho(x_{i},y_{j-1},z_k)}{y_{j+1}-y_{j-1}},\ldots \right.\\ &\left. \quad \frac{\rho(x_{i},y_j,z_{k+1})-\rho(x_{i},y_j,z_{k-1})}{z_{k+1}-z_{k-1}} \right)
\end{align*}.

The volume integral of $f$ is derived in the similar manner:
\begin{equation}\label{volume}
    \int_{\Omega_m} f(x,y,z)\; \mathrm{d}S =\frac{1}{2}\left( \sum_{(i,j,k)\in I_1} f(x_i,y_j,z_k)h^3 + \sum_{(i,j,k)\in I_1\cup I_o} f(x_i,y_j,z_k)h^3\right),
\end{equation}
Here $I_1$ contains all the grid points inside $\Omega_m$ and $I_o$ is the set of the irregular grid points defined at the surface area estimation equation \eqref{surface-area}. The desired volume of an enclosed molecular surface is attained by setting $f(x,y,z)=1$.




\subsection{Machine learning strategy with EISA}
\begin{figure}[!ht]
    \centering
     \includegraphics[width=1.0\textwidth]{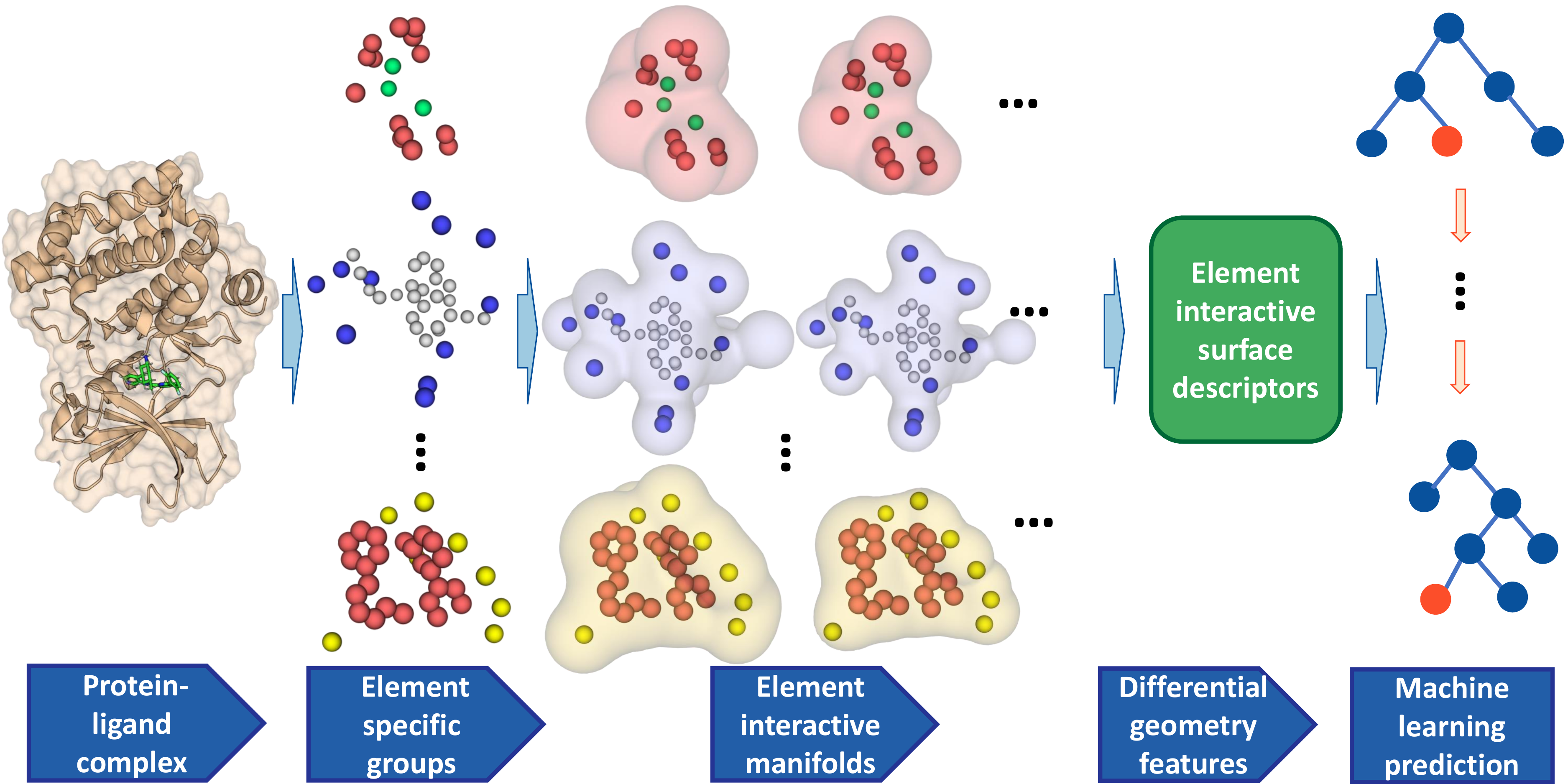}
    \caption{ Illustration of EISA learning strategy using a complex with PDBID: 5dwr (first column).  The second column represents the element-specific groups for Carbon-Fluoride, Nitrogen-Hydrogen, and Oxygen-Carbon from top to bottom, respectively. The corresponding element interactive manifolds are shown in the third column. Different manifolds are generated by varying the isovalue $0<c<1$. The fourth column presents the surface area-based descriptors obtained from various manifolds. In the final column, the advanced machine learning models such as the gradient boosting trees integrate these differential geometry features for training and prediction.}
    \label{fig:flowchart}
\end{figure}
The descriptors of the element interactive surface area (EISA) for a molecule or molecular complex provide robustness and scalable features for machine learning or deep learning-based models to learn the diverse biomolecular datasets.  The global and local element interactive densities, respectively defined in \eqref{eq:global_element_specific_density} and \eqref{eq:local_element_specific_density}, give rise to the corresponding global and local surface area descriptors. Furthermore, by varying the isovalue $c$ for attaining the isosurface $\Gamma_{kk'}=\{(x,y,z)\in \Omega:\rho_{kk'}=c\}$ of the element interactive manifold, one can arrive at multiple surfaces for a given molecule at different resolutions. That enables us to capture molecular surfaces at various scales, which embed the physical and chemical interactions between protein and ligand atoms at different ranges. The learning strategy with EISA descriptors are summarized in Fig. \ref{fig:flowchart}.

The EISA representations are ready to be integrated with wide variety of machine learning algorithms such as support vector machine \cite{cang2015topological}, random forest \cite{nguyen2017rigidity}, gradient boosting trees \cite{nguyen2019agl}, artificial neural networks \cite{gao20202d}, and convolutional neural networks \cite{cang2018representability}. However, we only use gradient boosting trees (GBTs) in this work instead of optimizing machine learning algorithm selections. 
We use GBT module in scikit-learn v0.24.1 package with the following parameters: $\texttt{n\_estimators}=10 000$, $\texttt{max\_depth} = 7$, \texttt{min\_samples\_split} = 3, \texttt{learning\_rate} = 0.01, \texttt{loss} = ls, \texttt{subsample} = 0.3, and \texttt{max\_features} = sqrt. These
parameter values are selected from the extensive tests on PDBbind datasets and are uniformly used in all our validation
tasks in this work.

\section{Results}
In this section, we demonstrate the performance of the proposed element interactive surface area (EISA) strategy for protein-ligand binding affinity prediction from three standard benchmarks in drug design. 

\subsection{Model parametrization}
For convenience, we use the notation $\mathrm{EISA}_{\kappa,\tau}^{\alpha}$ to indicate the element interactive surface areas (EISAs) generated by using kernel type $\alpha$ and corresponding kernel parameters $\kappa$ and $\tau$. Here, $\alpha=E$ and $\alpha=L$ refer to the generalized exponential and generalized Lorentz kernels, respectively. 
And $\tau$ is used such that $\eta_{kk'}=\tau (\bar{r}_k + \bar{r}_{k'})$, where $\bar{r}_k$ and
$\bar{r}_{k'}$ are the van der Waals radii of element type $k$ and element type $k'$ , respectively. Kernel parameters $\kappa$ and $\tau$ are selected
based on the cross validation with a random split of the training data.
We propose a EISA representation in which multiple kernels are parametrized at different scale ($\eta$) values. In this work, we
consider at most two kernels. As a straightforward notation extension, two kernels can be parametrized by $\mathrm{EISA}_{\kappa_1,\tau_1;\kappa_2,\tau_2}^{\alpha_1,\alpha_2}$.
Each of these kernels gives rise to one set of features. Since there are two ways of formulating the interactive surface areas, global surface (see Eq. (\ref{eq:global_element_specific_density}) and local surface (see Eq. \ref{eq:local_element_specific_density}), we finalize our notation to demonstrate two different kinds of surface calculation: $^{\text{glo}}\mathrm{EISA}_{\kappa_1,\tau_1;\kappa_2,\tau_2}^{\alpha_1,\alpha_2}$ and $^{\text{loc}}\mathrm{EISA}_{\kappa_1,\tau_1;\kappa_2,\tau_2}^{\alpha_1,\alpha_2}$. While the first notation stands for the global interactive surface, the latter indicates the local surface area.

\subsection{Datasets}
We are interested in using our EISA method to predict the binding affinities of protein-ligand complexes. A standard benchmark for such a prediction is the PDBbind database. Three popular PDBbind datasets, namely CASF--2007, CASF--2013, and CASF--2016, are employed to test the performance of our method. Each PDBbind dataset has a hierarchical structure consisting of the following subsets: a general set, a refined set, and a core set. The latter set is a subset of the previous one. The PDBbind database provides 3D coordinates of ligands and
their receptors obtained from experimental measurement via Protein Data Bank. In each benchmark, it is standard to use the refined set, excluding the core set, as a training set to build a predictive model for the
binding affinities of the complexes in the test set (i.e., the core set). More information about these datasets is offered on the PDBbind website \url{http://pdbbind.org.cn/}. A summary of the dataset is provided in Table \ref{tab:PDBbind_dataset}.

\begin{table}[!htbp]
\begin{center}
\caption{Summary of PDBbind datasets used in the present work\Bstrut}

\begin{tabular}{l c c}
	\hline
	Dataset & Training set complexes & Test set complexes\Tstrut\Bstrut\\
	\hline
    CASF--2007 benchmark & 1105 & 195\Tstrut\Bstrut\\
    CASF--2013 benchmark & 3516 & 195\Tstrut\Bstrut\\
    CASF--2016 benchmark & 3772 & 285\Tstrut\Bstrut\\
    \hline
\end{tabular}
\label{tab:PDBbind_dataset}
\end{center}
\end{table}

\subsection{Model performance and discussion}
\subsubsection{Hyperparameters and model setting}
To achieve the optimal EISA-Score's performances on each benchmark, we carefully optimize its hyperparameters on each training set. These hyperamaters include kernel parameters $\tau\in [0.5,6]$ and  $\kappa \in [0.5,10]$ with an increment of 0.5 and higher values in $\{15,20\}$. Moreover, the cutoff distance $d_c$ between 5 \AA \, and 12 \AA\, with an increment of 1. See Table \ref{tab:PDBbind_parameter_domain} for the summary of the hyperparameters' domain.  The element interactive surface area is described by four commonly occurring atom types, {C, N, O, S}, in protein and 10 commonly atom types, {H, C, N, O, F, P, S, Cl, Br, I}, in ligands. Note that we only employ the generalized exponential kernel in the current work since the generalized Lorentz kernel yield the similar accuracy \cite{nguyen2017rigidity,nguyen2019dg}.

 In global surface models, $^\text{glo}\text{EISA}$, with a given set of kernel parameters $(\tau, \kappa)$, cutoff distance $d_c$, and a pair of element types $\mathcal{T}_k$ and $\mathcal{T}_{k'}$, we consider 16 isovalues $c$ in the interval $[0.05,\dots, 0.8]$ with an increment of 0.05. That results in 16 surface area values. We then achieve 6 descriptors by taking the sum, mean, median, maximum, minimum, and standard deviation of those area values. Furthermore, there are $4\times 10=40$ combinations between protein and ligand element types. Finally, we encode the binding interaction in a protein-ligand complex into a vector of fixed length at $6\times40=240$ components.
 
 In local surface models, $^\text{loc}\text{EISA}$, besides using a similar hyperparameter setting of the global approach, we only select one single isovalue $c$ in $[0.1, 0.75]$ for element types $\mathcal{T}_k$ and $\mathcal{T}_{k'}$. However, a different atomic position from the element type $\mathcal{T}_k$ will generate a different interactive manifold resulting in a different surface area. To get a scalable representation, we calculate the sum, mean, median, maximum, minimum, and standard deviation of those various values. With 40 possible combinations betwee protein and ligand atom types, one can attain a descriptor of size 240 for a given complex.
 
\begin{table}[!htb]
\begin{center}
\caption{The ranges of EISA-Score's hyperparameters\Bstrut}

\begin{tabular}{c c}
	\hline
	Parameter & Domain\Tstrut\Bstrut\\
	\hline
	 $\tau$  & $\{0.5, 1.0, \dots, 6\}$ \Tstrut\Bstrut\\
     $\kappa$ & $\{0.5, 1, \dots, 10\} \cup \{15, 20\}$\Tstrut\Bstrut\\
     $d_c$ & $\{5, 6, \dots, 12\}$\Tstrut\Bstrut\\
     $c$ & $\{0.05, 0.1, \dots, 0.8\}$\Tstrut\Bstrut\\

    \hline
\end{tabular}
\label{tab:PDBbind_parameter_domain}
\end{center}
\end{table}

\subsubsection{Results and discussion}
There are several hyperparameters of our proposed models, $^\text{glo}\text{EISA}$, need to be carefully optimized for each benchmark. For a sake of achieving fairness performances, we only use the training data set to carry on the grid search on the designated domains mentioned in Table \ref{tab:PDBbind_parameter_domain} via the results of the cross validation (CV) tests. We execute 20 CV runs for each hyperparameter set, and the criteria are based on the best median Pearson's correlation coefficient $R_p$.

\paragraph{CASF--2007}

\begin{figure}[!htb]
\begin{center}
\includegraphics[width=1\linewidth]{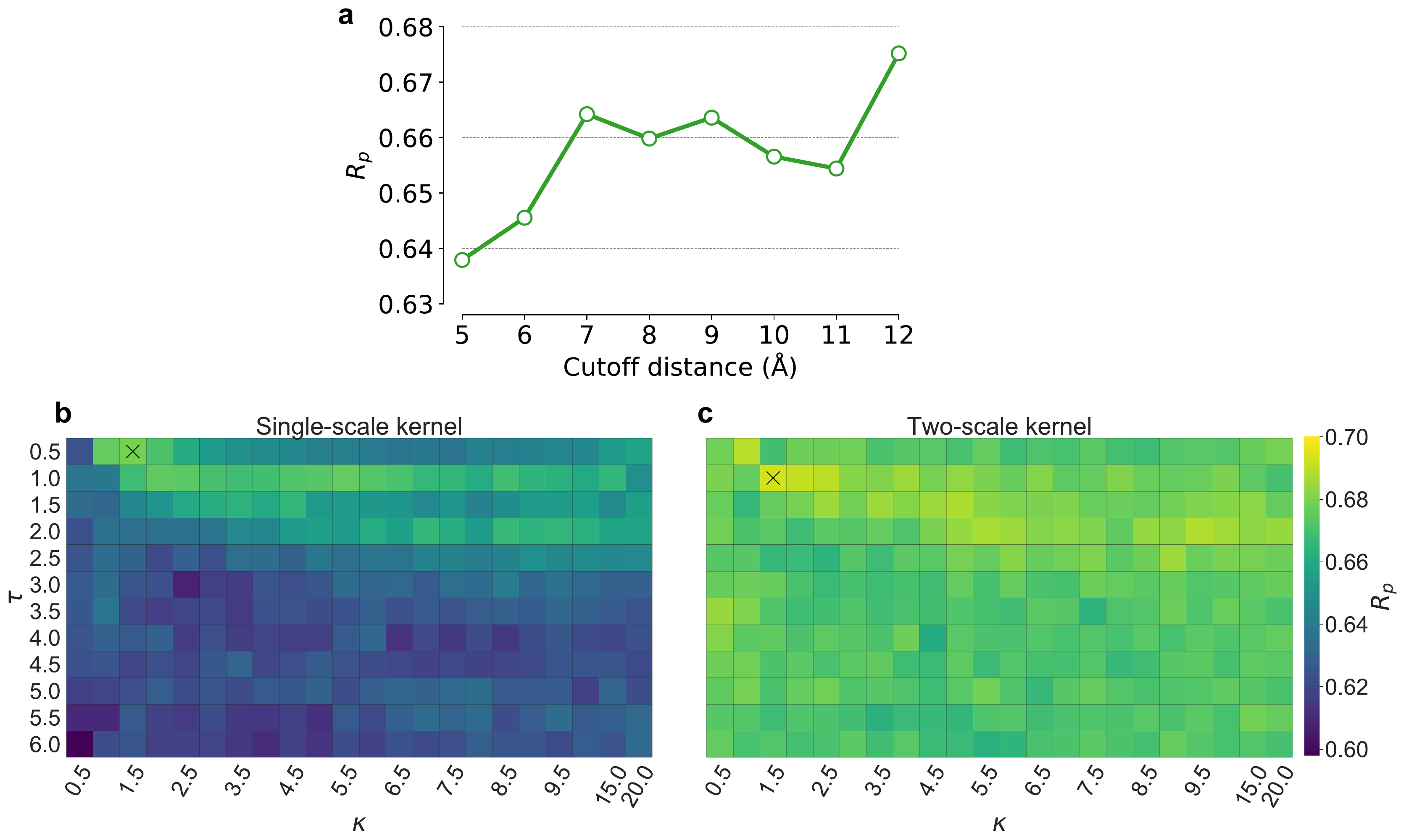}
\caption{Optimized parameters for the global surface model on CASF--2007 dataset. a) Global surface five-fold CV results for various cutoff distances on CASF--2007 training set. We fix  the kernel parameters $(\kappa, \tau) = (2,1)$. For a given cutoff distance value, we generate 16 surface areas based on 16 different isovalues in $[0.05,\dots,0.8]$. Global surface five-fold CV results for CASF--2007 training set with a) single-scale exponential kernel and b) two-scale exponential kernel. The best parameter locations are marked by ``x''. Specifically, the best parameters for single-scale kernel model are $(\kappa, \tau) = (1.5,0.5)$ with the corresponding Pearson's correlation coefficient $R_p=0.678$. Second optimized kernel parameters are  $(\kappa, \tau) = (1.5,1)$ producing $R_p=0.693$.
}
\label{fig:v2007_global_surface_optimized_parameters}
\end{center}
\end{figure}

\begin{figure}[!htb]
\begin{center}
\includegraphics[width=1\linewidth]{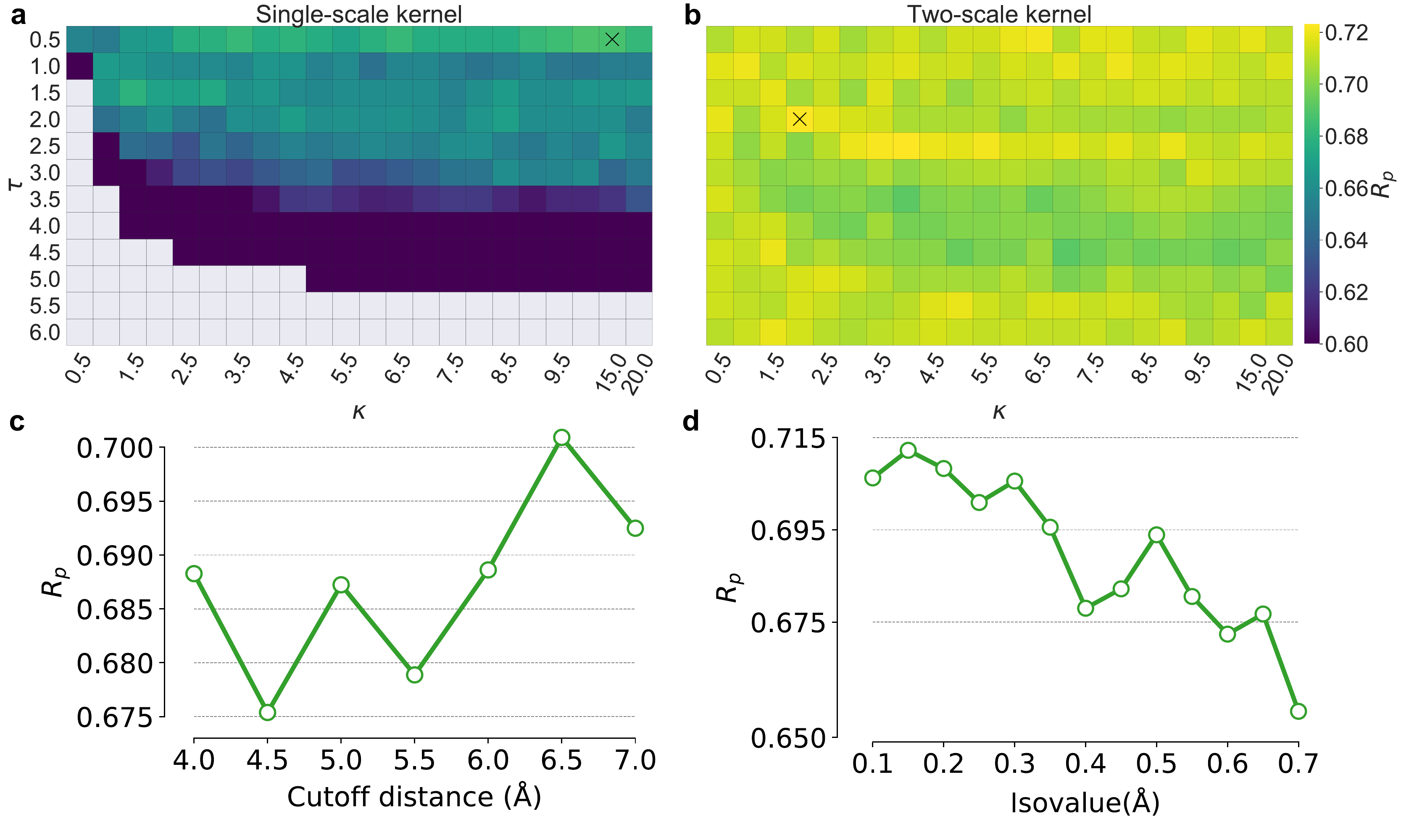}
\caption{Optimized parameters for the local surface model on CASF--2007 dataset. a) Local surface five-fold CV results for single-scale scenario and it is found that the best kernel parameters $(\kappa, \tau) = (15, 0.5)$ and corresponding median $R_p=0.688$. Note that there are empty values in the Figure a) since $R_p$ cannot be determined at the choice of $\kappa$ and $\tau$ b) Five-fold CV results for single-scale approach and the best second kernel parameters are $(\kappa, \tau) = (2, 2)$ producing the best $R_p=0.726$. The marker ``x'' indicates the position having the best $R_p$. c) The five-fold CV results of the local surface model with respect to the cutoff distance $d_c$. The best cutoff distance is $d_c=6.5$\AA, and $R_p=0.701$. d) The five-fold CV results of the local surface model when the isovalue $c$ varies from 0.1 to 0.7. Optimal isovalue $c$ is found to be 0.15 and the corresponding $R_p$ is 0.712.
}
\label{fig:v2007_local_surface_optimized_parameters}
\end{center}
\end{figure}

\begin{table}[!htbp]
\begin{center}
\caption{Performance of various EISA models on the CASF--2007 test set\Bstrut}

\begin{tabular}{l l c}
	\hline
	Model & $R_p$ & RMSE (kcal/mol)\Tstrut\Bstrut\\
	\hline
	\multicolumn{3}{c}{Results of Global Surface Model}\Tstrut\Bstrut\\
	\hline
		$^\text{glo}\mathrm{EISA}_{1.5,0.5}^{E,12}$ & 0.801 & 2.01\Tstrut\Bstrut\\
		$^\text{glo}\mathrm{EISA}_{1.5,0.5;1.5,1}^{EE,12}$ & 0.807 & 2.00\Tstrut\Bstrut\\
    \hline
    \multicolumn{3}{c}{Results with Local Surface}\Tstrut\Bstrut\\
    \hline
    $^\text{loc}\mathrm{EISA}_{15,0.5}^{E;6.5;0.15}$ & 0.807 & 1.986\Tstrut\Bstrut\\
    $^\text{loc}\mathrm{EISA}_{15,0.5;2,2}^{EE;6.5;0.15}$ & 0.793 & 2.046\Tstrut\Bstrut\\
    \hline
    \multicolumn{3}{c}{Results with Consensus Method}\Tstrut\Bstrut\\
    \hline
    Consensus\{$^\text{glo}\mathrm{EISA}_{1.5,0.5}^{E,12}$, $^\text{loc}\mathrm{EISA}_{15,0.5}^{E;6.5;0.15}$\} & {\bf 0.825} & {\bf 1.941}\Tstrut\Bstrut\\
    Consensus\{$^\text{glo}\mathrm{EISA}_{1.5,0.5;1.5,1}^{EE,12}$, $^\text{loc}\mathrm{EISA}_{15,0.5;2,2}^{EE;6.5;0.15}$\} & 0.817 &1.984\Tstrut\Bstrut\\
    \hline
\end{tabular}
\label{tab:CASF2007_results}
\end{center}
\end{table}

\begin{figure}[!htbp]
\begin{center}
\includegraphics[width=1\linewidth]{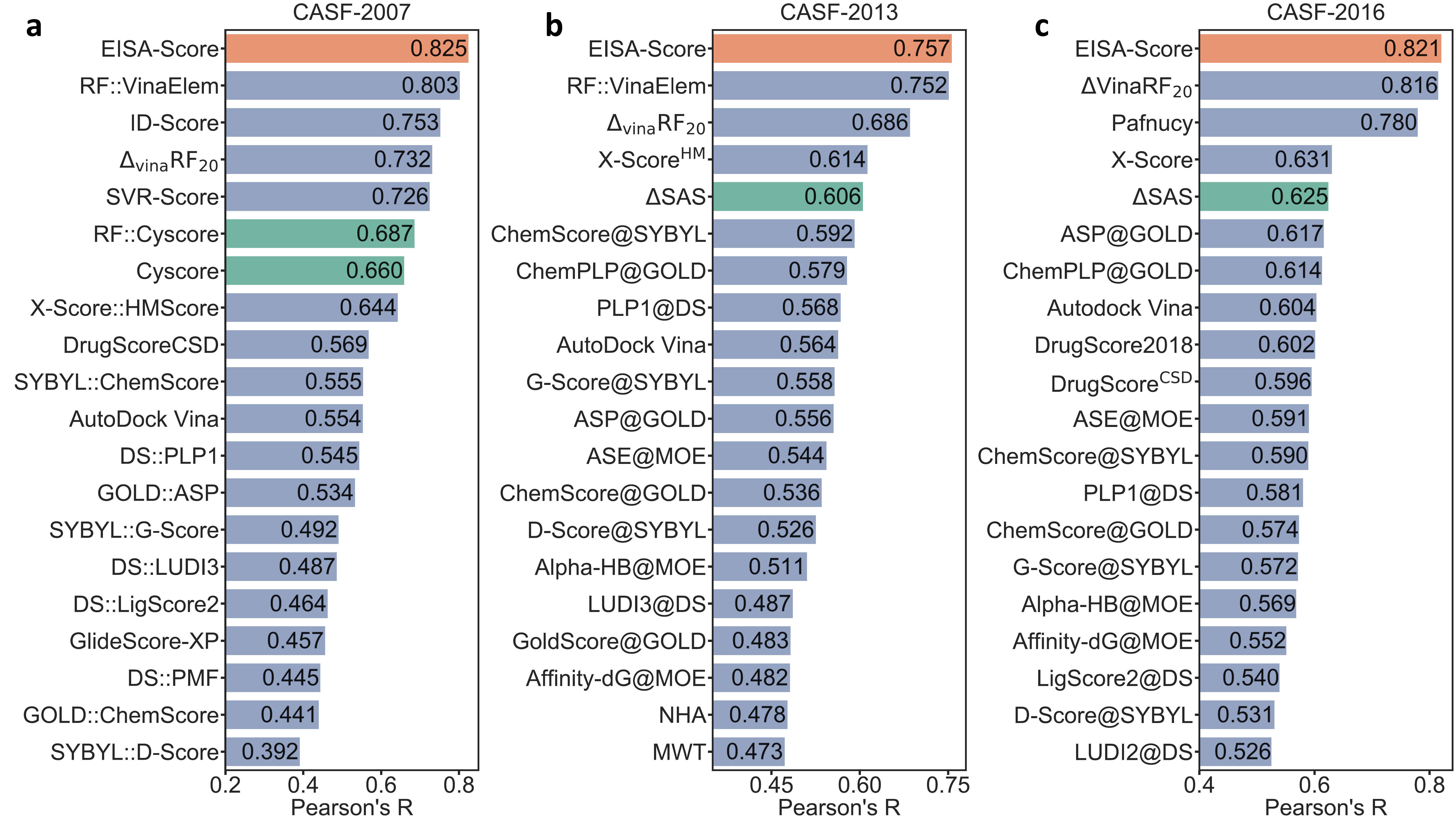}
\caption{Performance comparison different scoring functions on CASF benchmarks. Our proposed model in this work, EISA-Score, is highlighted in red, the other geometric based scoring functions are highlighted in green, and the rest is in purple. a) CASF--2007:  the performances of other methods taken from previous studies \cite{cheng2009comparative,ballester2010machine,li2013id,li2015improving,li2014substituting,cao2014improved,wang2017improving}. Our EISA-Score achieves $R_p$=0.825 and RMSE=1.941 kcal/mol. b) CASF--2013: the other results are extracted from \cite{wang2017improving,li2014comparative,li2015improving}. Our EISA-Score achieves $R_p$=0.757 and RMSE=2.113 kcal/mol. c) CASF--2016: our EISA-Score achieves $R_p$=0.821 and RMSE = 1.835 kcal/mol, other scoring functions are discussed in \cite{su2018comparative,stepniewska2018development,wang2017improving}.}
\label{fig:CASF_scoring_power}
\end{center}
\end{figure}

At first, we carry out the five-fold CV of the global surface models $^\text{glo}\text{EISA}$ on the CASF--2007 training data. To explore the optimal cutoff distance values $d_c$, we fix the kernel parameters $(\kappa, \tau) = (2,1)$, select 16 isovalues $c$ in$[0.05,\dots,0.8]$, but vary the $d_c$ between  5 \AA \, and 12 \AA\, with increment of 1. Figure \ref{fig:v2007_global_surface_optimized_parameters}a reveals that $d_c=12$ \AA\, yields the best median Pearson's correlation coefficient $R_p$. 

We now explore the optimal exponential kernel parameters for a single-scale model $^\text{glo}\mathrm{EISA}_{\kappa,\tau}^{\mathrm{E}, 12}$ where  $\tau\in [0.5,6]$ and  $\kappa \in [0.5,10]$ with  increment of 0.5. We also consider high values of $\kappa \in \{15,20\}$.  Figure \ref{fig:v2007_global_surface_optimized_parameters}b plots all the CV results and shows that $(\kappa, \tau) = (1.5, 1)$ gives the best median $R_p=0.678$ for the global surface model. The two-scale kernel model, $^\text{glo}\text{EISA}^{\mathrm{E}\mathrm{E}, 12}_{1.5,0.5,\kappa_2,\tau_2}$, is built on top of the previously optimized single scale . The optimal second kernel parameters $(\kappa_2,\tau_2)$ is explored via CV experiments and the result of each parameter combination is illustrated in Figure \ref{fig:v2007_global_surface_optimized_parameters}c. We found that $^\text{glo}\text{EISA}^{\mathrm{E}, 12}_{1.5,0.5, 1.5, 1}$ produces the best median $R_p=0.693$ on the CASF--2007 training set. It is interesting to observe that the single-kernel model $^\text{glo}\text{EISA}^{\mathrm{E}\mathrm{E},12}_{1.5,0.5,\kappa_2,\tau_2}$ performs well on the 195 complexes from the test set of the CASF--2007 benchmark with the reported $R_p=0.801$  and the root-mean-square error (RMSE) = 2.01 kcal/mol. While the two-scale model $^\text{glo}\text{EISA}^{\mathrm{E}\mathrm{E},12}_{1.5,0.5, 1.5, 1}$ performs slightly better than its predecessor and achieves $R_p=0.807$ and RMSE = 2.00 kcal/mol. Those results are reported in Table \ref{tab:CASF2007_results}.

The second kind of our surface-based model is the local surface based approach, $^\text{loc}\text{EISA}$, that measures the various different surface areas between a single protein atom and all the ligand atoms. There is a slightly difference in term of the parameter choice between the global and local models. While the global surface areas utilize various isovalues between 0.05 and 0.8, the local surface approach will explore the isovalue to generate the best surface model. But at first, while we fix the isovalue $c=0.25$, and cutoff distance $d_c=5$ \AA, we vary the kernel parameters $\kappa$ and $\tau$ in their designated domains (see Table \ref{tab:PDBbind_parameter_domain}). Figure \ref{fig:v2007_local_surface_optimized_parameters}a visualizes that CV test and reports the best kernel parameters $(\kappa=15, \tau=0.5)$ with $R_p=0.688$. In the next step, we investigate the best cutoff distance $d_c$ for the local surface based model with previously optimized single-kernel parameters ($\kappa=15$ and $\tau=0.5)$ and an isovalue $c=0.25$. In this experiment, we vary $d_c$ between 4 \AA, and 7 \AA, with increment of 0.5, then we find out the optimal cutoff distance is 6.5 \AA, that produces the median $R_p=0.701$ on the 5-fold CV of CASF--2007 training set, see Figure \ref{fig:v2007_local_surface_optimized_parameters}c.

The isovalue $c$ is the next parameter we would like to optimize for our local surface model,  $^\text{loc}\text{EISA}^{\mathrm{E}, 6.5, c}_{15,0.5}$. We search $c$ in the discrete domain between 0.1 and 0.75 with increment of 0.5. Figure \ref{fig:v2007_local_surface_optimized_parameters}d reveals that using isovalue $c=0.15$ will be the best choice for $^\text{loc}\text{EISA}^{\mathrm{E}, 6.5, c}_{15,0.5}$ with the reported median $R_p=0.712$. Similar to the global surface model, we are interested in extending the single-scale EISA-score $^\text{loc}\text{EISA}^{\mathrm{E}, 6.5, 0.15}_{15,0.5}$ to the two-scale one $^\text{loc}\text{EISA}^{\mathrm{E}, 6.5, 0.15}_{15,0.5,\kappa_2,\tau_2}$. Figure \ref{fig:v2007_local_surface_optimized_parameters}b summarizes the performances of the current model on the 5-fold experiments with respect to different values of $\kappa_2$ and $\tau_2$. And we conclude that  $(\kappa_2, \tau_2) =(2,2)$ gives us that optimal two-scale learner achieving the best median $R_p=0.726$.

Table \ref{tab:CASF2007_results} report the efficiency of one-kernel and two-kernel local surface models on the CASF--2007 test set. Interestingly, with only one-single scale, the local surface model, $^{\text{loc}}\mathrm{EISA}_{15,0.5}^{E;6.5;0.15}$ performs similarly to the two-scale approach using the global surface features. Its $R_p$ value is 0.807 but its RMSE is as low as 1.986 kcal/mol and is lower than of the global surface model. Unfortunately, the two-kernel version of the local EISA-score does not improve the what one-kernel has already achieved. In fact,the $R_p$ of $^\text{loc}\mathrm{EISA}_{15,0.5;2,2}^{EE;6.5;0.15}$ is just 0.793 and the corresponding RMSE is 2.046 kcal/mol. The consensus model which is the aggregation of the predicted values from unrelated models is acclaimed to often improve the overall performance \cite{nguyen2017rigidity,nguyen2019dg,nguyen2019dg}. For that reason, we include the consensus version in our proposed models. As seen from Table \ref{tab:CASF2007_results}, the consensus approach formed by single scale between the global and local surface areas, Consensus\{$^\text{glo}\mathrm{EISA}_{1.5,0.5}^{E,12}$, $^\text{loc}\mathrm{EISA}_{15,0.5}^{E;6.5;0.15}$\},  gives rise to the best one with $R_p=0.825$ and RMSE = 1.941 kcal/mol. While the consensus of the two-scales models produce the second best $R_p$ at 0.817 and RMSE at 1.984 kcal/mol. 

In addition, we compare the scoring power of our proposed EISA-Score against the state-of-the-art scoring functions in the literature \cite{cheng2009comparative,ballester2010machine,li2013id,li2015improving,li2014substituting,cao2014improved}. Figure \ref{fig:CASF_scoring_power}a plots the aforementioned comparison and clearly the dominance of our EISA model in the scoring power task.  Note that the geometrical-based models, Cyscore \cite{cao2014improved} and RF::Cyscore \cite{li2014substituting}, are highlighted in the green color. Specifically, Cyscore used area and curvature dependent descriptors. However, its performance ($R_p=0.660$) is not as good as our proposed EISA-Score ($R_p=0.825$) due to the lack of the examination of pairwise element types inducing interactive manifolds. Furthermore, one can cite another reason is the missing machine learning power in the Cyscore model. However, Li and his colleague \cite{li2014substituting} solved that concern by replacing the Cyscore's original scoring function by the random forest, and the result is not promising with the reported $R_p$ as low as 0.687. These results confirm the efficiency and robustness of the proposed element specific surface area based descriptors for protein-ligand complexes.

\begin{figure}[!htbp]
\begin{center}
\includegraphics[width=1\linewidth]{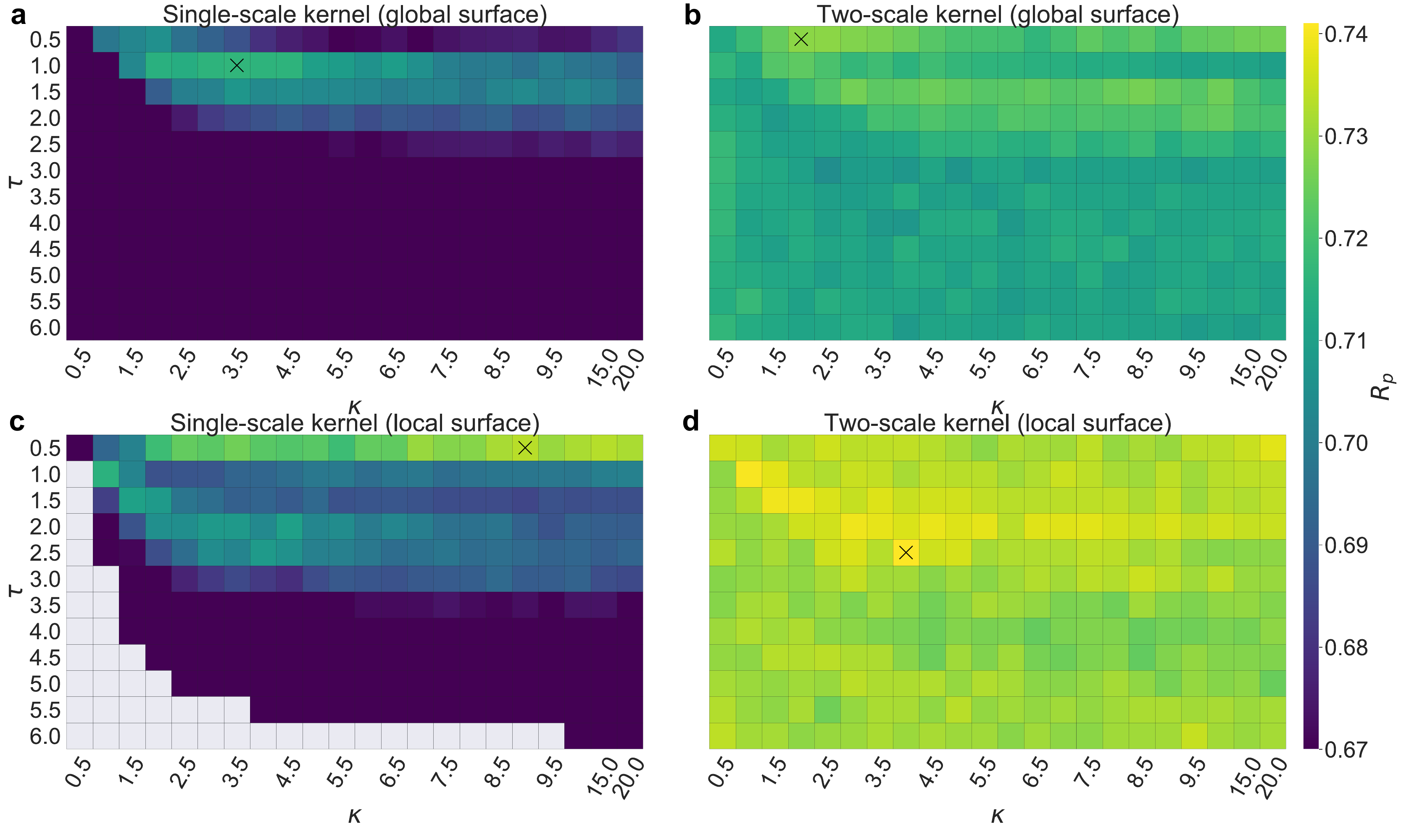}
\caption{Optimized parameters of the global and local surface models for CASF--2013.  Note that the marker ``x'' indicates the position having the best $R_p$ and  the empty values in the Figures due to the fact that $R_p$ cannot be determined at the choice of $\kappa$ and $\tau$. a) Single-scale global surface and its optimal kernel parameters $(\kappa, \tau) = (3.5, 1)$ and corresponding median $R_p=0717$. b) Two-scale kernel global surface and its optimal parameters for the second kernel $(\kappa, \tau) = (2, 0.5)$ and corresponding median $R_p=0729$. c) Single-scale local surface model and its optimal parameters for the second kernel $(\kappa, \tau) = (3, 1)$ and corresponding median $R_p=0.715$. d) Two-scale local surface model and its optimal parameters for the second kernel $(\kappa, \tau) = (3, 0.5)$ and corresponding median $R_p=0.727$.
}
\label{fig:v2013_optimized_parameters}
\end{center}
\end{figure}

\paragraph{CASF--2013} In this second benchmark among the CASF family, we carry out the similar hyperparamters optimization strategy to the CASF--2007 approach. For the simplicity, we use the optimized cutoff distance $d_c=12$ \AA, found from CASF--2007 dataset for the global surface model. To explore the most optimal parameters for the first kernel, we again perform 5-fold CV on CASF--2013 training data and find out that $(\kappa_1=3.5, \tau_1=1)$ gives the best $R_p=0.717$ (see Figure \ref{fig:v2013_optimized_parameters}a). To construct the second kernel, we simply fix the first kernel parameters, and vary the second kernel parameter in the interested domain (see Table \ref{tab:PDBbind_parameter_domain}). We found the best parameter for the second kernel $\kappa_2=2$ and $\tau_2=0.5$ with best median $R_p=0.729$ (see Figure \ref{fig:v2013_optimized_parameters}b). Finally, we achieve the optimal one-kernel global surface model 	$^\text{glo}\mathrm{EISA}_{3.5,1}^{E,12}$ and the optimal two-kernels global surface model 
		$^\text{glo}\mathrm{EISA}_{3.5,1;2,0.5}^{EE,12}$. These models are utilized to predict the unseen complexes from the test set of CASF--2013. As seen from Table \ref{tab:CASF2013_results}, the performances of single kernel and two kernels, respectively, achieve ($R_p$=0.684, RMSE=2.286 kcal/mol) and ($R_p=0.724$, RMSE = 2.180 kcal/mol). There is a considerable improvement from the single kernel to two kernels model in compassion to the what we have observed in CASF--2007. The size of the training set (1105 for CASF--2007 and 3516 for CASF--2013) can play a huge factor role in our multi-scale strategy.

		To reduce the search time cost of hyperparameters for the local surface approach, we use the optimized cutoff distance $d_c=6.5$ \AA, and the isovalue $c=0.15$ which are explored from the CASF--2007 experiment. These parameters are pretty consistent among different protein-ligand complexes. Therefore, we speculate there is little room for improvement if we re-optimize those parameters. Similar to the global surface scheme, we first search for the optimal one-kernel model. Figure \ref{fig:v2013_optimized_parameters}c plots the 5-fold CV results of $\mathrm{EISA}_{\kappa_1,\tau_1}^{E;6.5;0.15}$ on the training set of CASF--2013, and we conclude that $k_1=0$ and $\tau_1=0.5$ will yield the best $R_p=0.734$. Again, for the two-kernel model $\mathrm{EISA}_{9,0.5;\kappa_2,\tau_2}^{EE;6.5;0.15}$, we use the optimized value from the single-scale model for the first kernel, and explore the optimal ones for the second kernel. As see in Figure \ref{fig:v2013_optimized_parameters}d, $\kappa_2=4$ and $\tau_2=2.5$ produces the best $R_p=0.741$. Finally, we evaluate the scoring power of two selected local surface models, $^\text{local}\mathrm{EISA}_{9,0.5}^{E;6.5;0.15}$  and $^\text{loc}\mathrm{EISA}_{9,0.5;4,2.5}^{EE;6.5;0.15}$, on the CASF--2013 test set. It is comparable to what we observed in CASF--2007, the one-scale local surface model ($R_p$=0.749) performs a bit better than its counterpart ($R_p$ = 0.741), albeit a bigger training data. Our optimal strategy still relies on the consensus design where the consensus between two two-scale models,  Consensus\{$^\text{glo}\mathrm{EISA}_{3.5,1;2,0.5}^{EE,12}$,	$^\text{loc}\mathrm{EISA}_{9,0.5;4,2.5}^{EE;6.5;0.15}$\}, delivers the best $R_p$ as high as 0.756 and the corresponding RMSE = 2.113 kcal/mol. See Table \ref{tab:CASF2013_results} for the completion of results. Our EISA-Score again tops other published models on CASF--2013 as indicated in Figure \ref{fig:CASF_scoring_power}b. It is worth mentioning that, we also include other surface area-based model, $\Delta\mathrm{SAS}$ \cite{li2014comparative}, which used the solvent-accessible surface area of the buried ligand molecule when forming the complex. However, $\Delta\mathrm{SAS}$'s performance is not promising with $R_p$ as low as 0.606 due to the lack of the greater details of the buried surface for specific element types.

\begin{table}[!htbp]
\begin{center}
\caption{Performance of various EISA models on the CASF--2013 test set\Bstrut}

\begin{tabular}{l l l}
	\hline
	Model & $R_p$ & RMSE (kcal/mol)\Tstrut\Bstrut\\
	\hline
	\multicolumn{3}{c}{Results with global Surface}\Tstrut\Bstrut\\
	\hline
		$^\text{glo}\mathrm{EISA}_{3.5,1}^{E,12}$ & 0.684 & 2.286\Tstrut\Bstrut\\
		$^\text{glo}\mathrm{EISA}_{3.5,1;2,0.5}^{EE,12}$ & 0.724 &2.180\Tstrut\Bstrut\\
    \hline
    \multicolumn{3}{c}{Results with local Surface}\Tstrut\Bstrut\\
    \hline
    $^\text{loc}\mathrm{EISA}_{9,0.5}^{E;6.5;0.15}$ & 0.749 & {\bf 2.102}\Tstrut\Bstrut\\
    $^\text{loc}\mathrm{EISA}_{9,0.5;4,2.5}^{EE;6.5;0.15}$ & 0.741 &2.129\Tstrut\Bstrut\\
    \hline
    \multicolumn{3}{c}{Results with Consensus Method}\Tstrut\Bstrut\\
    \hline
    Consensus\{$^\text{glo}\mathrm{EISA}_{3.5,1}^{E,12}$, $^\text{loc}\mathrm{EISA}_{9,0.5}^{E;6.5;0.15}$\} & 0.741 & 2.155\Tstrut\Bstrut\\
    Consensus\{$^\text{glo}\mathrm{EISA}_{3.5,1;2,0.5}^{EE,12}$,	$^\text{loc}\mathrm{EISA}_{9,0.5;4,2.5}^{EE;6.5;0.15}$\} & {\bf 0.756} &2.113\Tstrut\Bstrut\\
    \hline
\end{tabular}
\label{tab:CASF2013_results}
\end{center}
\end{table}

\begin{figure}[!htbp]
\begin{center}
\includegraphics[width=1\linewidth]{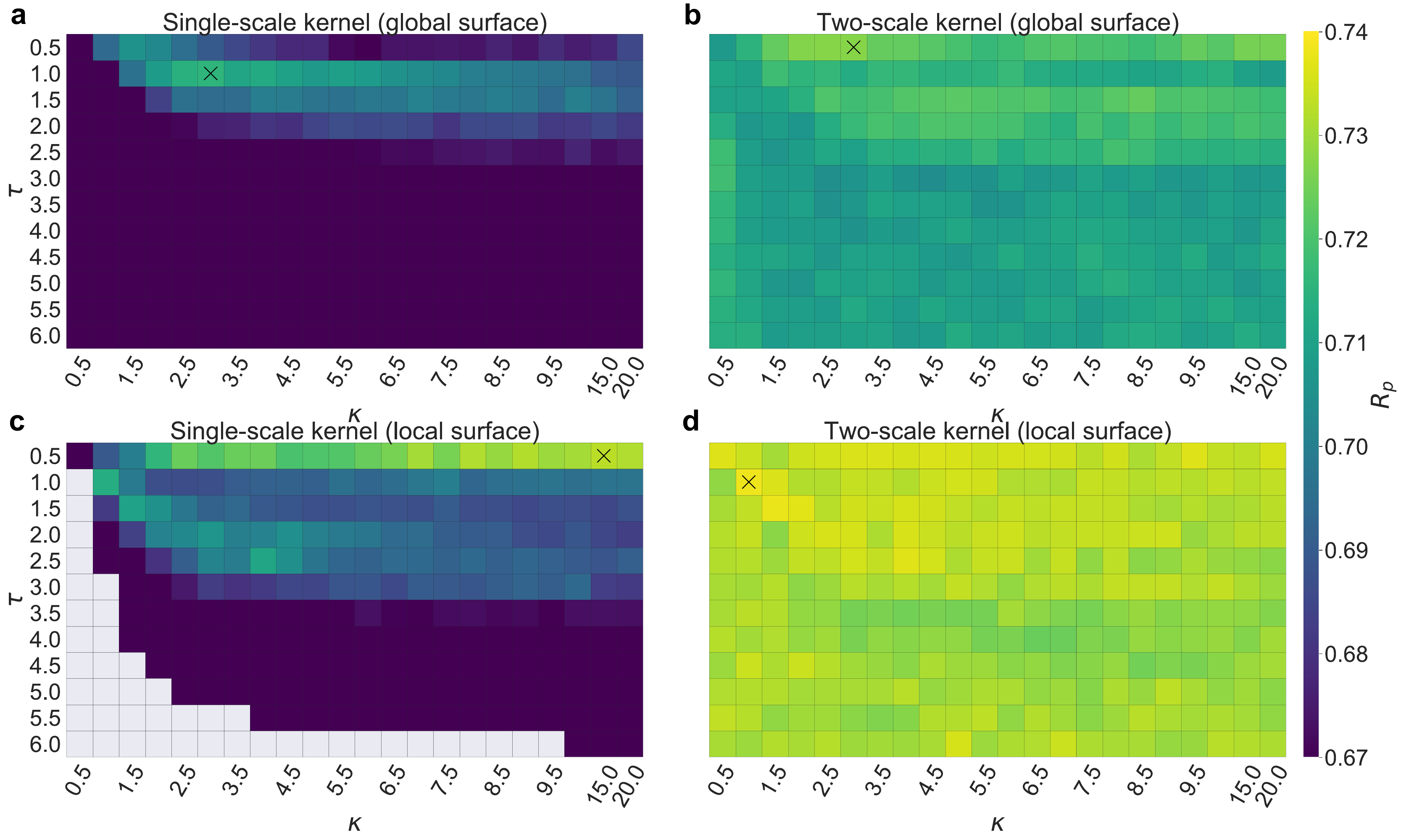}
\caption{Optimized parameters of the global and local surface models for CASF--2016.  Note that the marker ``x'' indicates the position having the best $R_p$ and  the empty values in the Figures due to the fact that $R_p$ cannot be determined at the choice of $\kappa$ and $\tau$. a) Single-scale global surface and its optimal kernel parameters $(\kappa, \tau) = (3, 1)$ and corresponding median $R_p=0.715$. b) Two-scale kernel global surface and its optimal parameters for the second kernel $(\kappa, \tau) = (3, 0.5)$ and corresponding median $R_p=0.727$. c) Single-scale local surface model and its optimal parameters for the second kernel $(\kappa, \tau) = (15, 0.5)$ and corresponding median $R_p=0.733$. d) Two-scale local surface model and its optimal parameters for the second kernel $(\kappa, \tau) = (1, 1)$ and corresponding median $R_p=0.738$.
}
\label{fig:v2016_optimized_parameters}
\end{center}
\end{figure}

\paragraph{CASF--2016} For this final benchmark, we perform the hyperparmameters search alike to what we proposed for CASF--2013. The global surface design will use the ideal distance cutoff $d_c=12$ \AA\, achieved from CASF--2007 experiments. The optimal single-scale model for CASF--2016 is found to be $^\text{glo}\mathrm{EISA}_{3,1}^{E,12}$, where its $R_p$ from the five-fold CV on 3772 complexes of CASF--2016 training set is equal to 0.715. On top this single-scale model, the two scales continues improve the CV performances with its bets model being as $^\text{glo}\mathrm{EISA}_{3,1;3,0.5}^{EE,12}$ and its $R_p$=0.727. Figures \ref{fig:v2016_optimized_parameters}a and \ref{fig:v2016_optimized_parameters}b summarize the CV results for various kernel parameters combinations.
		
The local surface approach uses the optimal isovalue $c=0.15$ and distance cutoff $d_c=6.5$ \AA\, realized from CASF--2007 five-fold results. Figures \ref{fig:v2016_optimized_parameters}c and \ref{fig:v2016_optimized_parameters}d reports the CV results on CASF--2016 with respect to the single-scale and two-scale parameter choices. Specifically,   $^\text{loc}\mathrm{EISA}_{15,0.5}^{E;6.5;0.15}$ is the best single-scale representative with the median 5-fold CV $R_p$ = 0.727. Furthermore, the best two-scale candidate is found to be $^\text{loc}\mathrm{EISA}_{15,0.5;1,1}^{EE;6.5;0.15}$ with the corresponding median 5-fold CV $R_p$ = 0.738.

Lastly, the aforementioned desirable EISA models are trained on the training data of CASF--2016 and are utilized to predict the binding energies of 285 complexes in CASF--2016 test set. Table \ref{tab:CASF2013_results} lists the results of these models including the consensus strategies. The familiar trend has been observed here. Two-scale models, $^\text{glo}\mathrm{EISA}_{3,1;3,0.5}^{EE,12}$ and    $^\text{loc}\mathrm{EISA}_{15,0.5;1,1}^{EE;6.5;0.15}$,  bring about the most outstanding performance among the non-consensus ones. In addition, the consensus models improve the existing methods. Specifically, Consensus\{$^\text{glo}\mathrm{EISA}_{3,1;3,0.5}^{EE,12},$	$^\text{loc}\mathrm{EISA}_{15,0.5;1,1}^{EE;6.5;0.15}$\} reaches the $R_p$=0.821 on the test set while its stand-alone models $^\text{glo}\mathrm{EISA}_{3,1;3,0.5}^{EE,12}$ and $^\text{loc}\mathrm{EISA}_{15,0.5;1,1}^{EE;6.5;0.15}$ scores $R_p$=0.798 and $R_p$=0.795, respectively. CASF--2016 is a prevalent benchmark which attract numerous scoring functions relying on it to test their scoring power \cite{su2018comparative,stepniewska2018development,wang2017improving}. As seen in Figure \ref{fig:CASF_scoring_power}c, it is encouraging to see our EISA-Score outperforming other state-of-the-art methods. It is noted that, among other 20 scoring functions listed in Figure \ref{fig:CASF_scoring_power}c, only $\Delta \text{SAS}$ \cite{su2018comparative} solely leans on the the surface descriptors. However its performance on CASF--2016 is unfavorable with $R_p$ =0.625 as opposed to 0.821 of our proposed EISA-Score. This result again confirms the rigorous and robust capacity of our novel surface area-based descriptors for drug design.

\begin{table}[!htbp]
\begin{center}
\caption{Performance of various EISA models on the CASF--2016 test set\Bstrut}

\begin{tabular}{l l l}
	\hline
	Model & $R_p$ & RMSE (kcal/mol)\Tstrut\Bstrut\\
	\hline
	\multicolumn{3}{c}{Results with Global Surface}\Tstrut\Bstrut\\
	\hline
		$^\text{glo}\mathrm{EISA}_{3,1}^{E,12}$ & 0.769 & 1.989\Tstrut\Bstrut\\
		$^\text{glo}\mathrm{EISA}_{3,1;3,0.5}^{EE,12}$ & 0.798 &1.888\Tstrut\Bstrut\\
    \hline
    \multicolumn{3}{c}{Results with Local Surface}\Tstrut\Bstrut\\
    \hline
    $^\text{loc}\mathrm{EISA}_{15,0.5}^{E;6.5;0.15}$ & 0.791 & 1.883\Tstrut\Bstrut\\
    $^\text{loc}\mathrm{EISA}_{15,0.5;1,1}^{EE;6.5;0.15}$ & 0.795 &1.881\Tstrut\Bstrut\\
    \hline
    \multicolumn{3}{c}{Results with Consensus Method}\Tstrut\Bstrut\\
    \hline
    Consensus\{$^\text{glo}\mathrm{EISA}_{3,1}^{E,12}, ^\text{loc}\mathrm{EISA}_{15,0.5}^{E;6.5;0.15}$\} & 0.813 & 1.873\Tstrut\Bstrut\\
    Consensus\{$^\text{glo}\mathrm{EISA}_{3,1;3,0.5}^{EE,12},	^\text{loc}\mathrm{EISA}_{15,0.5;1,1}^{EE;6.5;0.15}$\} & {\bf 0.821} & {\bf 1.835}\Tstrut\Bstrut\\
    \hline
\end{tabular}
\label{tab:Pdbbind2016_results_new}
\end{center}
\end{table}
\newpage
\section{Conclusion}
The molecular surface representations are well-known for the biological structure modeling to reveal the biomolecular properties and activities. However, their relationship to the biological functions is often encoded in the realm of the physical models such as Poisson-Boltzmann equation and Poisson-Nernst-Planck model. Unfortunately, the problematic parameter choices of these physical models have overshadowed the valuable information extracted from the molecular surface. There are some recent efforts to directly incorporate the surface area descriptors to capture the protein-ligand potency \cite{cao2014improved,dong2021prediction}. However, the conventional surface area models do not portray crucial physical and chemical interactions such as non-covalent bonds, hydrogen bonds, van der Waals interactions, etc., which lead to discouraging results and limited capacity to handle diverse biomolecular datasets. These issues call for robustness and scalable  surface area representations for biomolecular structures.

This work proposes a novel element interactive surface area score (EISA-Score) for protein-ligand binding prediction and can be extended to handle drug-related problems. Our proposed models construct scalable element interactive manifolds instead of a single surface representation for a whole complex often used in the standard approaches. The innovative surface areas help encode the physical and biological information mentioned above, which have been missed in conventional methods. Our EISA-Score offers two types of surface area models, namely global and local surface. Specifically, while the global surface area strategy provides the overall molecular representation between protein and ligand atoms, the local approach focuses on describing the local manifold formed by a specific protein atom and ligand molecule. Our molecular surfaces are induced by the discrete-to-continuum mapping powered by the correlation function such as exponential and Lorentz kernels. 

Due to the high sensitivity of the hyperparameters, including isovalue, kernel power, and kernel scalar factor in our surface generation, we carefully perform the cross validation on the training data to select the optimal surface descriptors for the protein-ligand complexes. As a result, our proposed EISA-Score achieves superior performances over state-of-the-art methods on three mainstream benchmarks, namely CASF--2007 \cite{cheng2009comparative}, CASF--2013 \cite{li2014comparative}, and CASF--2016 \cite{su2018comparative}. These encouraging results confirm our surface-area-based models' robustness, reliability, and accuracy in the binding affinity prediction for small molecules, which is an essential task in drug design.

\section*{Conflict of interest}
The authors declare that they have no conflict of interest.

\section*{Availability}
\vspace{-0.15cm}
The source code is available at Github: \url{https://github.com/NguyenLabUKY/EISA-Score}.

 \section*{Acknowledgements}
This work was supported in part by NSF Grants  
DMS-2053284, DMS-2151802, and University of Kentucky Startup Fund.  \\

\bibliographystyle{unsrt}
\bibliography{refs}

\begin{thebibliography}{10}

\bibitem{corey1953molecular}
Robert~B Corey and Linus Pauling.
\newblock Molecular models of amino acids, peptides, and proteins.
\newblock {\em Review of Scientific Instruments}, 24(8):621--627, 1953.

\bibitem{koltun1965precision}
Walter~L Koltun.
\newblock Precision space-filling atomic models.
\newblock {\em Biopolymers: Original Research on Biomolecules}, 3(6):665--679,
  1965.

\bibitem{lee1971interpretation}
Byungkook Lee and Frederic~M Richards.
\newblock The interpretation of protein structures: estimation of static
  accessibility.
\newblock {\em Journal of molecular biology}, 55(3):379--IN4, 1971.

\bibitem{richards1977areas}
Frederic~M Richards.
\newblock Areas, volumes, packing, and protein structure.
\newblock {\em Annual review of biophysics and bioengineering}, 6(1):151--176,
  1977.

\bibitem{connolly1985depth}
Michael~L Connolly.
\newblock Depth-buffer algorithms for molecular modelling.
\newblock {\em Journal of Molecular Graphics}, 3(1):19--24, 1985.

\bibitem{grant2001smooth}
J~Andrew Grant, Barry~T Pickup, and Anthony Nicholls.
\newblock A smooth permittivity function for {Poisson--Boltzmann} solvation
  methods.
\newblock {\em Journal of computational chemistry}, 22(6):608--640, 2001.

\bibitem{vorobjev1997sims}
Yury~N Vorobjev and Jan Hermans.
\newblock {SIMS}: computation of a smooth invariant molecular surface.
\newblock {\em Biophysical Journal}, 73(2):722--732, 1997.

\bibitem{sanner1996reduced}
Michel~F Sanner, Arthur~J Olson, and Jean-Claude Spehner.
\newblock Reduced surface: an efficient way to compute molecular surfaces.
\newblock {\em Biopolymers}, 38(3):305--320, 1996.

\bibitem{grant2007gaussian}
JA~Grant, BT~Pickup, MJ~Sykes, CA~Kitchen, and A~Nicholls.
\newblock The {Gaussian Generalized Born} model: application to small
  molecules.
\newblock {\em Physical Chemistry Chemical Physics}, 9(35):4913--4922, 2007.

\bibitem{li2013dielectric}
Lin Li, Chuan Li, Zhe Zhang, and Emil Alexov.
\newblock On the dielectric “constant” of proteins: smooth dielectric
  function for macromolecular modeling and its implementation in {DelPhi}.
\newblock {\em Journal of chemical theory and computation}, 9(4):2126--2136,
  2013.

\bibitem{wang2015pka}
Lin Wang, Lin Li, and Emil Alexov.
\newblock {pKa} predictions for proteins, {RNAs}, and {DNAs} with the
  {Gaussian} dielectric function using {DelPhi pKa}.
\newblock {\em Proteins: Structure, Function, and Bioinformatics},
  83(12):2186--2197, 2015.

\bibitem{petrey2003grasp2}
Donald Petrey and Barry Honig.
\newblock {GRASP2}: visualization, surface properties, and electrostatics of
  macromolecular structures and sequences.
\newblock In {\em Methods in enzymology}, volume 374, pages 492--509. Elsevier,
  2003.

\bibitem{chen2010kernel}
Wenyu Chen, Jianmin Zheng, and Yiyu Cai.
\newblock Kernel modeling for molecular surfaces using a uniform solution.
\newblock {\em Computer-Aided Design}, 42(4):267--278, 2010.

\bibitem{chan1998molecular}
Shek~Ling Chan and Enrico~O Purisima.
\newblock Molecular surface generation using marching tetrahedra.
\newblock {\em Journal of computational chemistry}, 19(11):1268--1277, 1998.

\bibitem{edelsbrunner1994three}
Herbert Edelsbrunner and Ernst~P M{\"u}cke.
\newblock Three-dimensional alpha shapes.
\newblock {\em ACM Transactions on Graphics (TOG)}, 13(1):43--72, 1994.

\bibitem{fraczkiewicz1998exact}
Robert Fraczkiewicz and Werner Braun.
\newblock Exact and efficient analytical calculation of the accessible surface
  areas and their gradients for macromolecules.
\newblock {\em Journal of computational chemistry}, 19(3):319--333, 1998.

\bibitem{hayryan2005new}
Shura Hayryan, Chin-Kun Hu, Jaroslav Sk{\v{r}}iv{\'a}nek, Edik Hayryane, and
  Imrich Pokorn{\`y}.
\newblock A new analytical method for computing solvent-accessible surface area
  of macromolecules and its gradients.
\newblock {\em Journal of computational chemistry}, 26(4):334--343, 2005.

\bibitem{liang1998analytical}
Jie Liang, Herbert Edelsbrunner, Ping Fu, Pamidighantam~V Sudhakar, and Shankar
  Subramaniam.
\newblock Analytical shape computation of macromolecules: I. molecular area and
  volume through alpha shape.
\newblock {\em Proteins: Structure, Function, and Bioinformatics}, 33(1):1--17,
  1998.

\bibitem{rychkov2007joint}
Georgy Rychkov and Michael Petukhov.
\newblock Joint neighbors approximation of macromolecular solvent accessible
  surface area.
\newblock {\em Journal of Computational Chemistry}, 28(12):1974--1989, 2007.

\bibitem{xu2009generating}
Dong Xu and Yang Zhang.
\newblock Generating triangulated macromolecular surfaces by euclidean distance
  transform.
\newblock {\em PloS one}, 4(12):e8140, 2009.

\bibitem{lange2020intrinsically}
Adrian~W Lange, John~M Herbert, Benjamin~J Albrecht, and Zhi-Qiang You.
\newblock Intrinsically smooth discretisation of connolly's solvent-excluded
  molecular surface.
\newblock {\em Molecular Physics}, 118(6):e1644384, 2020.

\bibitem{daberdaku2018computing}
Sebastian Daberdaku and Carlo Ferrari.
\newblock Computing voxelised representations of macromolecular surfaces: A
  parallel approach.
\newblock {\em The International Journal of High Performance Computing
  Applications}, 32(3):407--432, 2018.

\bibitem{egan2018fast}
Raphael Egan and Fr{\'e}d{\'e}ric Gibou.
\newblock Fast and scalable algorithms for constructing solvent-excluded
  surfaces of large biomolecules.
\newblock {\em Journal of Computational Physics}, 374:91--120, 2018.

\bibitem{hermosilla2017interactive}
Pedro Hermosilla, Michael Krone, Victor Guallar, Pere-Pau V{\'a}zquez,
  {\`A}lvar Vinacua, and Timo Ropinski.
\newblock Interactive {GPU}-based generation of solvent-excluded surfaces.
\newblock {\em The Visual Computer}, 33(6):869--881, 2017.

\bibitem{liu2018efficient}
Tiantian Liu, Minxin Chen, and Benzhuo Lu.
\newblock Efficient and qualified mesh generation for gaussian molecular
  surface using adaptive partition and piecewise polynomial approximation.
\newblock {\em SIAM Journal on Scientific Computing}, 40(2):B507--B527, 2018.

\bibitem{liu2017eses}
Beibei Liu, Bao Wang, Rundong Zhao, Yiying Tong, and Guo-Wei Wei.
\newblock {ESES}: Software for e ulerian solvent excluded surface, 2017.

\bibitem{wei2005molecular}
GW~Wei, Yuhui Sun, YC~Zhou, and M~Feig.
\newblock Molecular multiresolution surfaces.
\newblock {\em arXiv preprint math-ph/0511001}, 2005.

\bibitem{bates2006minimal}
PW~Bates, GW~Wei, and Shan Zhao.
\newblock The minimal molecular surface.
\newblock {\em arXiv preprint q-bio/0610038}, 2006.

\bibitem{bates2008minimal}
Peter~W Bates, Guo-Wei Wei, and Shan Zhao.
\newblock Minimal molecular surfaces and their applications.
\newblock {\em Journal of Computational Chemistry}, 29(3):380--391, 2008.

\bibitem{bates2009geometric}
PW~Bates, Zhan Chen, Yuhui Sun, Guo-Wei Wei, and Shan Zhao.
\newblock Geometric and potential driving formation and evolution of
  biomolecular surfaces.
\newblock {\em Journal of mathematical biology}, 59(2):193--231, 2009.

\bibitem{wei2012variational}
Guo-Wei Wei, Qiong Zheng, Zhan Chen, and Kelin Xia.
\newblock Variational multiscale models for charge transport.
\newblock {\em siam REVIEW}, 54(4):699--754, 2012.

\bibitem{wei2013multiscale}
Guo-Wei Wei.
\newblock Multiscale, multiphysics and multidomain models i: Basic theory.
\newblock {\em Journal of Theoretical and Computational Chemistry},
  12(08):1341006, 2013.

\bibitem{chen2010differential}
Zhan Chen, Nathan~A Baker, and Guo-Wei Wei.
\newblock Differential geometry based solvation model {I}: {Eulerian}
  formulation.
\newblock {\em Journal of computational physics}, 229(22):8231--8258, 2010.

\bibitem{chen2011differential}
Zhan Chen, Nathan~A Baker, and Guo-Wei Wei.
\newblock Differential geometry based solvation model {II}: {Lagrangian}
  formulation.
\newblock {\em Journal of mathematical biology}, 63(6):1139--1200, 2011.

\bibitem{chen2011differential2}
Zhan Chen and Guo-Wei Wei.
\newblock Differential geometry based solvation model. {III}. {Quantum}
  formulation.
\newblock {\em The Journal of chemical physics}, 135(19):194108, 2011.

\bibitem{grant1995gaussian}
J~Andrew Grant and BT~Pickup.
\newblock A {Gaussian} description of molecular shape.
\newblock {\em The Journal of Physical Chemistry}, 99(11):3503--3510, 1995.

\bibitem{yu2008feature}
Zeyun Yu, Michael~J Holst, Yuhui Cheng, and J~Andrew McCammon.
\newblock Feature-preserving adaptive mesh generation for molecular shape
  modeling and simulation.
\newblock {\em Journal of Molecular Graphics and Modelling}, 26(8):1370--1380,
  2008.

\bibitem{mu2017geometric}
Lin Mu, Kelin Xia, and Guowei Wei.
\newblock Geometric and electrostatic modeling using molecular rigidity
  functions.
\newblock {\em Journal of Computational and Applied Mathematics}, 313:18--37,
  2017.

\bibitem{nguyen2017impact}
Duc~D Nguyen and Guo-Wei Wei.
\newblock The impact of surface area, volume, curvature, and {Lennard--Jones}
  potential to solvation modeling.
\newblock {\em Journal of computational chemistry}, 38(1):24--36, 2017.

\bibitem{chen2012variational}
Zhan Chen, Shan Zhao, Jaehun Chun, Dennis~G Thomas, Nathan~A Baker, Peter~W
  Bates, and GW~Wei.
\newblock Variational approach for nonpolar solvation analysis.
\newblock {\em The Journal of chemical physics}, 137(8):084101, 2012.

\bibitem{wang2015parameter}
Bao Wang and GW~Wei.
\newblock Parameter optimization in differential geometry based solvation
  models.
\newblock {\em The Journal of chemical physics}, 143(13):10B608\_1, 2015.

\bibitem{chen2012quantum}
Duan Chen, Zhan Chen, and Guo-Wei Wei.
\newblock Quantum dynamics in continuum for proton transport ii: Variational
  solvent--solute interface.
\newblock {\em International Journal for Numerical Methods in Biomedical
  Engineering}, 28(1):25--51, 2012.

\bibitem{chen2012quantum_2}
Duan Chen and Guo-Wei Wei.
\newblock Quantum dynamics in continuum for proton transport—generalized
  correlation.
\newblock {\em The Journal of Chemical Physics}, 136(13):04B606, 2012.

\bibitem{chen2013quantum}
Duan Chen and Guo-Wei Wei.
\newblock Quantum dynamics in continuum for proton transport i: Basic
  formulation.
\newblock {\em Communications in computational physics}, 13(1):285--324, 2013.

\bibitem{nguyen2017accurate}
Duc~D Nguyen, Bao Wang, and Guo-Wei Wei.
\newblock Accurate, robust, and reliable calculations of poisson--boltzmann
  binding energies.
\newblock {\em Journal of computational chemistry}, 38(13):941--948, 2017.

\bibitem{spolar1994coupling}
Ruth~S Spolar and M~Thomas Record~Jr.
\newblock Coupling of local folding to site-specific binding of proteins to
  dna.
\newblock {\em Science}, 263(5148):777--784, 1994.

\bibitem{crowley2005cation}
Peter~B Crowley and Adel Golovin.
\newblock Cation--$\pi$ interactions in protein--protein interfaces.
\newblock {\em Proteins: Structure, Function, and Bioinformatics},
  59(2):231--239, 2005.

\bibitem{dragan2004dna}
Anatoly~I Dragan, Christopher~M Read, Elena~N Makeyeva, Ekaterina~I Milgotina,
  Mair~EA Churchill, Colyn Crane-Robinson, and Peter~L Privalov.
\newblock Dna binding and bending by hmg boxes: energetic determinants of
  specificity.
\newblock {\em Journal of molecular biology}, 343(2):371--393, 2004.

\bibitem{sobolev1996molecular}
Vladimir Sobolev, Rebecca~C Wade, Gert Vriend, and Marvin Edelman.
\newblock Molecular docking using surface complementarity.
\newblock {\em Proteins: Structure, Function, and Bioinformatics},
  25(1):120--129, 1996.

\bibitem{das2009rapid}
Sourav Das, Arshad Kokardekar, and Curt~M Breneman.
\newblock Rapid comparison of protein binding site surfaces with property
  encoded shape distributions.
\newblock {\em Journal of chemical information and modeling},
  49(12):2863--2872, 2009.

\bibitem{geng2011multiscale}
Weihua Geng and Guo-Wei Wei.
\newblock Multiscale molecular dynamics using the matched interface and
  boundary method.
\newblock {\em Journal of computational physics}, 230(2):435--457, 2011.

\bibitem{baker2005improving}
Nathan~A Baker.
\newblock Improving implicit solvent simulations: a poisson-centric view.
\newblock {\em Current opinion in structural biology}, 15(2):137--143, 2005.

\bibitem{chen2011mibpb}
Duan Chen, Zhan Chen, Changjun Chen, Weihua Geng, and Guo-Wei Wei.
\newblock Mibpb: a software package for electrostatic analysis.
\newblock {\em Journal of computational chemistry}, 32(4):756--770, 2011.

\bibitem{zheng2011second}
Qiong Zheng, Duan Chen, and Guo-Wei Wei.
\newblock Second-order poisson--nernst--planck solver for ion transport.
\newblock {\em Journal of computational physics}, 230(13):5239--5262, 2011.

\bibitem{zheng2011poisson}
Qiong Zheng and Guo-Wei Wei.
\newblock Poisson--boltzmann--nernst--planck model.
\newblock {\em The Journal of chemical physics}, 134(19):194101, 2011.

\bibitem{bergstrom2003absorption}
Christel~AS Bergstr{\"o}m, Melissa Strafford, Lucia Lazorova, Alex Avdeef,
  Kristina Luthman, and Per Artursson.
\newblock Absorption classification of oral drugs based on molecular surface
  properties.
\newblock {\em Journal of medicinal chemistry}, 46(4):558--570, 2003.

\bibitem{nguyen2019dg}
Duc~Duy Nguyen and Guo-Wei Wei.
\newblock {DG-GL:} differential geometry-based geometric learning of molecular
  datasets.
\newblock {\em International journal for numerical methods in biomedical
  engineering}, 35(3):e3179, 2019.

\bibitem{cao2014improved}
Yang Cao and Lei Li.
\newblock Improved protein--ligand binding affinity prediction by using a
  curvature-dependent surface-area model.
\newblock {\em Bioinformatics}, 30(12):1674--1680, 2014.

\bibitem{dong2021prediction}
Lina Dong, Xiaoyang Qu, Yuan Zhao, and Binju Wang.
\newblock Prediction of binding free energy of protein--ligand complexes with a
  hybrid molecular mechanics/generalized born surface area and machine learning
  method.
\newblock {\em ACS omega}, 6(48):32938--32947, 2021.

\bibitem{xia2013multiscale}
Kelin Xia, Kristopher Opron, and Guo-Wei Wei.
\newblock Multiscale multiphysics and multidomain models—flexibility and
  rigidity.
\newblock {\em The Journal of chemical physics}, 139(19):11B614\_1, 2013.

\bibitem{xia2015multiresolution}
Kelin Xia, Zhixiong Zhao, and Guo-Wei Wei.
\newblock Multiresolution persistent homology for excessively large
  biomolecular datasets.
\newblock {\em The Journal of chemical physics}, 143(13):10B603\_1, 2015.

\bibitem{xia2016review}
Kelin Xia and Guo-Wei Wei.
\newblock A review of geometric, topological and graph theory apparatuses for
  the modeling and analysis of biomolecular data.
\newblock {\em arXiv preprint arXiv:1612.01735}, 2016.

\bibitem{cheng2009comparative}
Tiejun Cheng, Xun Li, Yan Li, Zhihai Liu, and Renxiao Wang.
\newblock Comparative assessment of scoring functions on a diverse test set.
\newblock {\em Journal of chemical information and modeling}, 49(4):1079--1093,
  2009.

\bibitem{li2014comparative}
Yan Li, Li~Han, Zhihai Liu, and Renxiao Wang.
\newblock Comparative assessment of scoring functions on an updated benchmark:
  2. evaluation methods and general results.
\newblock {\em Journal of chemical information and modeling}, 54(6):1717--1736,
  2014.

\bibitem{su2018comparative}
Minyi Su, Qifan Yang, Yu~Du, Guoqin Feng, Zhihai Liu, Yan Li, and Renxiao Wang.
\newblock Comparative assessment of scoring functions: the casf-2016 update.
\newblock {\em Journal of chemical information and modeling}, 59(2):895--913,
  2018.

\bibitem{nguyen2017rigidity}
Duc~D Nguyen, Tian Xiao, Menglun Wang, and Guo-Wei Wei.
\newblock Rigidity strengthening: A mechanism for protein--ligand binding.
\newblock {\em Journal of chemical information and modeling}, 57(7):1715--1721,
  2017.

\bibitem{nguyen2019agl}
Duc~Duy Nguyen and Guo-Wei Wei.
\newblock Agl-score: algebraic graph learning score for protein--ligand binding
  scoring, ranking, docking, and screening.
\newblock {\em Journal of chemical information and modeling}, 59(7):3291--3304,
  2019.

\bibitem{cang2018representability}
Zixuan Cang, Lin Mu, and Guo-Wei Wei.
\newblock Representability of algebraic topology for biomolecules in machine
  learning based scoring and virtual screening.
\newblock {\em PLoS computational biology}, 14(1):e1005929, 2018.

\bibitem{opron2015communication}
Kristopher Opron, Kelin Xia, and Guo-Wei Wei.
\newblock Communication: Capturing protein multiscale thermal fluctuations.
\newblock {\em The Journal of chemical physics}, 142(21):06B401\_1, 2015.

\bibitem{smereka2006numerical}
Peter Smereka.
\newblock The numerical approximation of a delta function with application to
  level set methods.
\newblock {\em Journal of Computational Physics}, 211(1):77--90, 2006.

\bibitem{cang2015topological}
Zixuan Cang, Lin Mu, Kedi Wu, Kristopher Opron, Kelin Xia, and Guo-Wei Wei.
\newblock A topological approach for protein classification.
\newblock {\em Computational and Mathematical Biophysics}, 3(1), 2015.

\bibitem{gao20202d}
Kaifu Gao, Duc~Duy Nguyen, Vishnu Sresht, Alan~M Mathiowetz, Meihua Tu, and
  Guo-Wei Wei.
\newblock Are 2d fingerprints still valuable for drug discovery?
\newblock {\em Physical chemistry chemical physics}, 22(16):8373--8390, 2020.

\bibitem{ballester2010machine}
Pedro~J Ballester and John~BO Mitchell.
\newblock A machine learning approach to predicting protein--ligand binding
  affinity with applications to molecular docking.
\newblock {\em Bioinformatics}, 26(9):1169--1175, 2010.

\bibitem{li2013id}
Guo-Bo Li, Ling-Ling Yang, Wen-Jing Wang, Lin-Li Li, and Sheng-Yong Yang.
\newblock Id-score: a new empirical scoring function based on a comprehensive
  set of descriptors related to protein--ligand interactions.
\newblock {\em Journal of chemical information and modeling}, 53(3):592--600,
  2013.

\bibitem{li2015improving}
Hongjian Li, Kwong-Sak Leung, Man-Hon Wong, and Pedro~J Ballester.
\newblock Improving autodock vina using random forest: the growing accuracy of
  binding affinity prediction by the effective exploitation of larger data
  sets.
\newblock {\em Molecular informatics}, 34(2-3):115--126, 2015.

\bibitem{li2014substituting}
Hongjian Li, Kwong-Sak Leung, Man-Hon Wong, and Pedro~J Ballester.
\newblock Substituting random forest for multiple linear regression improves
  binding affinity prediction of scoring functions: Cyscore as a case study.
\newblock {\em BMC bioinformatics}, 15(1):1--12, 2014.

\bibitem{wang2017improving}
Cheng Wang and Yingkai Zhang.
\newblock Improving scoring-docking-screening powers of protein--ligand scoring
  functions using random forest.
\newblock {\em Journal of computational chemistry}, 38(3):169--177, 2017.

\bibitem{stepniewska2018development}
Marta~M Stepniewska-Dziubinska, Piotr Zielenkiewicz, and Pawel Siedlecki.
\newblock Development and evaluation of a deep learning model for
  protein-ligand binding affinity prediction.
\newblock {\em Bioinformatics}, 1:9, 2018.

\end{thebibliography}

\end{document}